\documentclass{aa}
\newcommand{\etal}{{\rm et al.~}}           
\newcommand{\NH}{\mbox{${\rm N_H}$ }}       
\newcommand{\NHunits}{\mbox{$~{\rm cm}^{-2}$} }   
\begin{document}

\title {A New Narrow-Line Seyfert~1 galaxy : RX~J1236.9+2656}

\author{ G. C. Dewangan\inst{1},  K. P. Singh\inst{1}, Paula Szkody\inst{2}$^,$\thanks{Based on observations with the Apache Point Observatory (APO) 3.5m telescope,
which is owned and operated by the Astrophysical Research Consortium (ARC).}, and D. W. Hoard\inst{3}$^{,\star}$
}

\institute{
Department of Astronomy \& Astrophysics, Tata Institute of Fundamental Research, Mumbai, 400~005, India
\and
Department of Astronomy, University of Washington, Seattle, WA 98195, USA 
\and 
Cerro Tololo Inter-American Observatory, Casilla 603, La Serena, Chile}

\thesaurus{11.09.1 RX~J1236.9+2656}
\maketitle
\markboth{Dewangan et al.:  A New Narrow-Line Seyfert~1 galaxy : RX~J1236.9+2656}{}
\begin{abstract}
We report identification of a narrow-line Seyfert 1 galaxy  
RX~J1236.9+2656. 
X-ray emission from the 
NLS1 galaxy undergoes long-term variability with 0.1--2.0~keV flux
changing by a factor of $\sim2$ within $\sim3{\rm~yr}$. 
The $ROSAT$ PSPC spectrum of 
RX~J1236.9+2656 is well represented by a power-law of $\Gamma_{X}=3.7_{-0.3}^{+0.5}$ 
absorbed by matter in our own Galaxy (\NH$ = 1.33\times10^{20}\NHunits$). Intrinsic 
soft X-ray luminosity of the
NLS1 galaxy is estimated to be $\sim 1.5\times10^{43}{\rm ~ erg~s^{-1}}$ in the 
energy band of $0.1-2.0{\rm~keV}$.
The optical spectrum of RX~J1236.9+2656 is typical of NLS1 galaxies and shows 
narrow Balmer emission lines ($1100{\rm~km~s^{-1}}<$ FWHM $<1700{\rm~km~s^{-1}}$) 
of H$\beta$, H$\alpha$, and forbidden lines of [O III] and [N II]. Fe II multiplets, 
usually present in optical spectra of NLS1 galaxies, are also detected in RX~J1236.9+2656. 

\end{abstract}

\section{INTRODUCTION}
Narrow-line Seyfert 1 (NLS1) galaxies are considered to be a special class of ``normal" Seyfert 1 galaxies because 
of their peculiar properties that distinguish them from the latter class. They are characterized 
by their optical spectra having permitted lines that are narrower than those found in the normal 
Seyfert 1 galaxies, e.g., full width at half maximum (FWHM) of H$\beta$ line is $\la2000{\rm~km~s^{-1}}$,  
relatively weak forbidden lines, $\frac{[O III]\lambda5007}{H\beta}<3$ (Osterbrock \& Pogge 1985), and  
strong Fe~II emission.
NLS1 galaxies also have distinctive soft X-ray properties as well. 
They show steep 
soft X-ray spectrum with little or no absorption above the Galactic value (Grupe \etal 1998). They often 
show rapid and large amplitude as well as long-term X-ray variability (Boller \etal 1993; 
Brandt, Pounds, \& Fink 1995; Grupe \etal 1995a,b). In spite of the dominance of soft X-ray emission, soft X-ray luminosity of NLS1 galaxies are similar to those of normal Seyfert 1s. $ASCA$ observations show that 
the hard X-ray ($2-10{\rm~keV}$) continua of NLS1s are also steeper than those of normal 
Seyfert 1s with broader H$\beta$ FWHM (Brandt, Mathur, \& Elvis 1997; Leighly 1999b).
  NLS1 galaxies also show more variability in hard X-rays than the normal Seyfert 1s 
(Leighly, 1999a). The spectral energy distribution (SED) from far-infrared (FIR) 
to X-rays of NLS1 galaxies appears to be similar to that of broad-line Seyfert 1 galaxies. However, the UV luminosity of NLS1 galaxies tends to be smaller than those of Seyfert 1s 
(Rodriguez-Pascual, Mas-Hesse, \& Santos-Lle\'{o} 1997).

Optical spectroscopy of Ultra-soft X-ray sources discovered with $Einstein$, and $ROSAT$ has been an efficient way to identify NLS1 galaxies (e.g, Puchnarewicz \etal 1992; Grupe \etal 1998).
As part of our programme to optically identify and 
study in detail the counterparts of the ultra-soft sources in the catalogue of 
Singh \etal (1995), we have discovered a NLS1 galaxy RX~J1236.9+2656.
The basic parameters of RX~J1236.9+2656 are given
in Table 1.
\begin{table}[h]
\caption{Basic parameters of RX~J1236.9+2656.}
\begin{tabular}{l}
\hline
Position$^1$ : $\alpha(J2000) = 12^{h}~36^{m}~57.0^{s}$ \\
~~~~~~~~~~~~~~ $\delta(J2000) = +26\degr~56\arcmin~50.0\arcsec$.  \\
Redshift$^1$ : $z$ = $0.225\pm.001$.  \\
Magnitude$^2$ : $B$ = $18.2$, $V$ = $17.1$. \\
\hline \\
$^1$ Present work \\
$^2$ US Naval Observatory (USNO) catalogue
\end{tabular}
\end{table}
Throughout this note, luminosities are calculated assuming an isotropic emission, a Hubble
constant of $H_{0}=75{\rm~km~s^{-1}~Mpc^{-1}}$ and a deceleration
parameter of $q_{0}=0$ unless otherwise specified.

\section{X-ray Observation, Analysis \& Results}
The region of the sky containing the source, RX~J1236.9+2656, was observed seven times with
the $ROSAT$ (Truemper et al. 1983) Position Sensitive Proportional
Counter (PSPC) during 1991--1993 and twice with the High Resolution Imager (HRI)
(Pfeffermann et al. 1987) in 1996 June--July. The exposure times were in the range $1420{\rm~s}-5422{\rm~s}$ for the PSPC observations while the two HRI observations were carried out with longer exposure times ($14545{\rm~s}$ and $16738{\rm~s}$). The offset of the source from the field center was $\sim17\arcmin$ for each of the PSPC and HRI observations. 

The X-ray source, RX~J1236.9+2656, was identified by overlaying the contours
of high resolution X-ray images obtained from $ROSAT$ HRI observations onto optical
images obtained from the Digital Sky Survey (DSS).
No other X-ray source, within the angular spread comparable 
to the point spread
function of $ROSAT$ HRI, was seen in the overlays.
Therefore,  X-ray emission from RX~J1236.9+2656 is not contaminated by
emission from any other source. HRI count rates for RX~J1236.9+2656 were obtained using a
circle of radius $50\arcsec$ for the source and an annulus of inner circle radius $60\arcsec$
and width $60\arcsec$ for background. The HRI count rates thus estimated are ($1.14\pm0.14)\times10^{-2}$ and ($1.23\pm0.13)\times10^{-2}{\rm~count~s^{-1}}$ for the two observations.
The spatial resolution with the PSPC at an offset of $16.5\arcmin$ is $\sim40\arcsec$ (half power radius) (Hasinger \etal 1993). Therefore,
PSPC count rates for RX~J1236.9+2656 were obtained using a circle of 
radius $2.5\arcmin$ for the source and 7 nearby circular regions of radii $2.25\arcmin$ 
for background. 
The PSPC count rates are ($3.53\pm0.46)\times10^{-2}$, ($3.19\pm0.57)\times10^{-2}$, ($3.19\pm0.60)\times10^{-2}$, ($3.54\pm0.57)\times10^{-2}$, ($2.23\pm0.87)\times10^{-2}$, ($3.26\pm0.34)\times10^{-2}$, and ($3.15\pm0.40)\times10^{-2}{\rm~count~s^{-1}}$ for the 7 PSPC observations.

In order to investigate the time variability of  soft X-ray emission from
RX~J1236.9+2656, we extracted light curves from the $ROSAT$ PSPC
observations using the same source and the background regions as described
 above and in the PSPC energy
band of 0.1--2.4 keV containing all the X-ray photons. The background
subtractions were carried out after appropriately scaling the background
light curves to have the same area as the source extraction area.
The light curves of RX~J1236.9+2656 do not show short-term variability during the PSPC observations. 
However, 
on a longer time scale of months to years, variability is clearly detected in the X-ray flux measurements plotted in Figure 2. 
The observed flux in the energy band $0.1-2.0{\rm ~keV}$, estimated from the best fit spectral model (see below), increased by about 
a factor of 2 within $\sim3{\rm~yr}$.  

{\begin{figure}
\vskip 6cm
\includegraphics{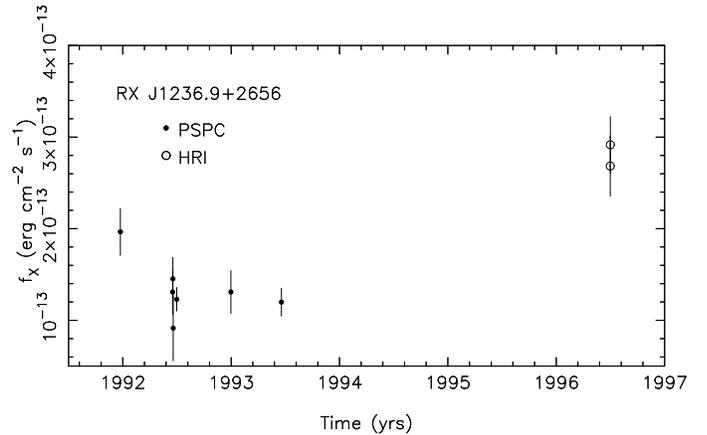}
\caption {Long-term $ROSAT$ light curve of RX~J1236.9+2656. The observed flux in the energy band of 0.1--2.0~keV has been plotted.}
\end{figure}}

For analyzing the X-ray spectra of RX~J1236.9+2656, we choose 3 PSPC
spectra corresponding to those observations
for which the exposure times were greater than $3000{\rm ~s}$. These observations were carried out on 1991 December 15, 1992 June 30, and 1993 June 17.
Photon energy spectra of RX~J1236.9+2656 were accumulated 
from their PSPC observations using the same source and background regions 
as stated above. $ROSAT$ PSPC pulse height data were appropriately 
re-grouped to improve the statistics.

We used the XSPEC spectral analysis package to
fit the data with spectral models. Appropriate ancillary  response file 
was used to account for the off-axis position of the source.
An appropriate response matrix
was used to define the energy response of
the PSPC.

Each of the 3 PSPC 
spectra was first fitted with a redshifted power-law model absorbed by an 
intervening medium with absorption cross-sections as
given by Balucinska-Church and McCammon (1992) and using the method of
$\chi^{2}$-minimization. 
The photon index ($\Gamma_{X}$) was found to be very steep in all the cases. 
It was found that the absorbing column density in each case is similar within 
errors to the Galactic value ($\NH=1.33\times10^{20}\NHunits$) measured from 
21-cm radio observations (Dickey \& Lockman 1990) along the direction of 
the source, indicating that all the X-ray
absorption is due to matter in our own Galaxy. Therefore, we have fitted the
power-law models to these spectra after fixing the neutral hydrogen column
density to the Galactic value. 
The best-fit minimum $\chi^{2}_{\nu}$ and the power-law
index do not change significantly from those obtained while varying the
\NH. The photon indices obtained for fixed \NH are, however,
better constrained. The values for $\Gamma_{X}$ are $3.0_{-0.4}^{+0.5}$, $4.2_{-0.7}^{+1.2}$, and $4.2_{-0.7}^{+2.0}$ for
the spectra observed on 1991 Dec 15, 1992 June 30, and 1993 June 17, respectively, and
are quite similar within the errors 
for all three spectra. The errors quoted, here and below,
were calculated at the $90\%$ confidence level based on  $\chi^2_{\rm min }$+2.71.
In order to better constrain the model parameters, 
we have fitted the above model to the three spectra jointly.  
The best-fit photon index is now $4.0_{-0.5}^{+0.5}$.
 The observed flux, based on the best-fit model parameters, 
is estimated to be $1.4\times10^{-13}{\rm ~erg~cm^{-2}~s^{-1}}$ in the energy band of
$0.1-2.0{\rm~keV}$. In the same energy band the intrinsic soft X-ray luminosity, corrected for Galactic absorption, 
of RX~J1236.9+2656  is calculated to be $1.6\times10^{43}{\rm~erg~s^{-1}}$.

We have also analyzed all the 7 PSPC 
spectra jointly. The best-fit phton index is $3.7^{+0.3}_{-0.5}$ which is similar to that obtained for the 3 PSPC observations above.
Thus, it is clear that all available PSPC spectral data of 
RX~J1236.9+2656 are 
well represented by a power-law of photon index $3.7_{-0.5}^{+0.3}$ absorbed 
by the matter in our own Galaxy. We have also fitted redshifted blackbody models, absorbed 
by an intervening medium, to the three PSPC spectra (exposure times $>$ 3000~s) of RX~J1236.9+2656 taken jointly as well as all the 7 data taken jointly. 
The absorbing column density derived from the model 
fit was lower than the Galactic value in that direction, indicating that the blackbody is not a suitable model. 
We then fixed the absorbing column to the Galactic value and carried out the joint 
fitting. The temperature thus obtained, $kT=79_{-24}^{+13}{\rm~eV}$, reflects the ultra-soft nature of RX~J1236.9+2656.

We have calculated the $ROSAT$ HRI flux of RX~J1236.9+2656 using the best-fit 
model parameters obtained from best-fit to all the 7 spectral data ($\Gamma_{X}=3.7$, $\NH=1.33\times10^{19}\NHunits$). 
The observed HRI fluxes are $2.7\times10^{-13}{\rm~erg~s^{-1}~cm^{-2}}$ 
and $2.9\times10^{-13}{\rm~erg~s^{-1}~cm^{-2}}$ in the energy band of $0.1-2.0\rm{~keV}$ for the two HRI observations. 
The HRI fluxes are about a factor of two higher than the values obtained from the PSPC observations.

\section{Optical Spectroscopy}
Low resolution optical spectroscopic observations of RX J1236.9+2656 were 
carried out at the 3.5m telescope of the Apache Point Observatory (APO) on
the nights of 1995 April 25, 30 and 2000 January 6. The integrations ranged
from 4-15 min, with the best spectrum obtained during the last observation.
The Double Imaging Spectrograph (DIS) was used in low resolution mode with a
$1.5\arcsec$ slit to obtain simultaneous blue and red spectra covering
${\rm 3800-5300\AA}$ in the blue and ${\rm 5600-9900\AA}$  in the red 
with a resolution of 14\AA.

{\begin{figure}
\vskip 7cm
\includegraphics{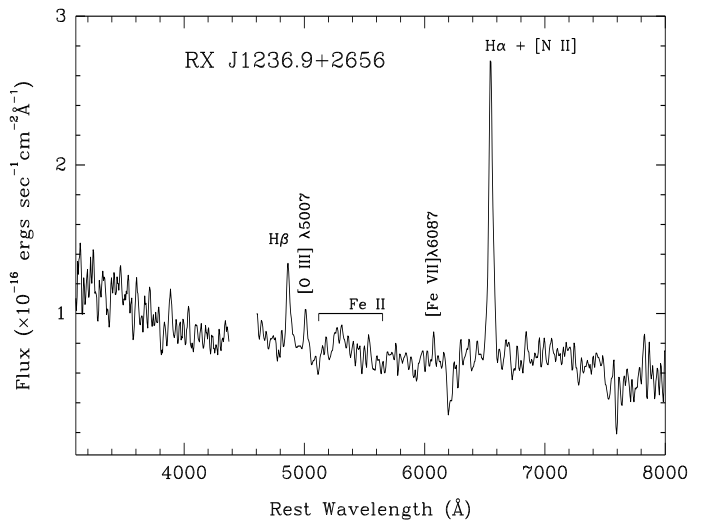}
\caption {Optical spectrum of RX~J1236.9+2656.
The gap around 4500\AA is due to the piece of the spectrum that was not
covered due to the dichroic beam which splits the light into the blue and
red spectrographs.}
\end{figure}}

The optical spectra were reduced using routines available within the IRAF
\footnote{{IRAF (Image Reduction and Analysis Facility) is distributed by
the National Optical Astronomy Observatories, which are operated by AURA, Inc.,
 under cooperative agreement with the National Science Foundation.}} software.
This included correcting the images using bias and flat fields, extracting
sky-subtracted one-dimensional spectra and calibrating the
wavelengths from HeNeAr lamps and the fluxes from standard stars.
In the spectrum of RX~J1236.9+2656, strong emission lines of Balmer H$\alpha$,
H$\beta$, and the forbidden line [O III]$\lambda5007$ are readily observed. Using the peak wavelengths and the rest wavelengths
of these lines, a redshift of 0.225$\pm0.001$ was derived from the 2000
January spectrum. The observed
spectrum was then corrected for this redshift and is shown in Figure 2.
The signal-to-noise 
ratio of the spectrum is $\sim7.6$ measured from the dispersion in the continuum 
region $6900{\rm~\AA}-7200{\rm~\AA}$. The continuum is observed
 to rise towards the blue end of the spectrum. 
Apart from the emission lines mentioned above, we have also identified Fe~II 
emission between 
$5070-5600{\rm~\AA}$ and a forbidden line of [Fe VII]$\lambda6087$ (see Fig. 2).  
We have fitted Gaussian 
profiles to the strong emission lines
using the profile fitting feature in the `splot' task within IRAF. Due to the poor 
resolution of the spectrum, it was not possible to deblend the H$\alpha$, and 
[N II]$\lambda\lambda6548,6584$ lines. A single Gaussian profile is 
not well fitted to the core of the H$\alpha$ line and the FWHM is overestimated 
because of the presence of [N II]$\lambda\lambda6548,6584$ lines. We  
also fitted a single Gaussian to the H$\alpha$ after excluding the wings.  
In this case, the FWHM can be considered as a lower limit.  
Thus we derive the FWHM of 
the H$\alpha$ line to be in the range of $1315-1692{\rm~km~s^{-1}}$. The widths have 
been corrected for instrumental broadening by subtracting, in quadrature, the 
instrumental broadening (FWHM = $14{\rm~\AA}$) from the observed FWHM. The H$\alpha$+[N II] 
flux is estimated to be $8.3\times10^{-15}{\rm~erg~cm^{-2}~s^{-1}}$ by integrating 
the flux over the H$\alpha$+[N II] profile.  
The profile of the Balmer line H$\beta$ is well fitted by a Lorentzian 
profile but poorly fitted with a Gaussian. The FWHMs of the best-fit Lorentzian and 
Gaussian profiles to the H$\beta$ line are $1122{\rm~km~s^{-1}}$ and 
$1392{\rm~km~s^{-1}}$, respectively. The observed widths of H$\alpha$ and H$\beta$, and the presence of Fe~II, [Fe~VII] emission are characteristic of NLS1 galaxies. 
The flux of the H$\beta$ 
line is estimated to be $2.0\times10^{-15}{\rm~erg~cm^{-2}~s^{-1}}$. Similarly the 
flux of the [O III]$\lambda5007$ line is calculated to be $7.6\times10^{-16}{\rm~erg~cm^{-2}~s^{-1}}$. Thus the ratio, $\frac{[O III]\lambda5007}{H\beta}$ is only $\sim0.38$. 

\section{Discussion}
RX~J1236.9+2656 is luminous in soft X-rays, with 
a rest frame intrinsic luminosity $\sim1.6\times10^{43}{\rm~erg~s^{-1}}$, in the energy band of 0.1--2.0~keV. Seyfert 1 galaxies studied by Rush \etal (1996) span over 4 orders of
magnitude in soft X-ray luminosity, from below $10^{42}{\rm~erg~s^{-1}}$
to above $10^{46}{\rm~erg~s^{-1}}$ in the energy band of 0.1--2.4~keV.
Thus, the X-ray luminosity of RX~J1236.9+2656
is similar to that of a Seyfert 1 galaxy. Assuming that the soft X-ray 
luminosity of RX~J1236.9+2656 is about $10\%$ of the bolometric luminosity, 
the lower limit to the mass of the central supermassive object or the 
Eddington mass is $\sim10^6{\rm~M\sun}$.

The galaxy RX~J1236.9+2656 shows long-term variability -- a change in intensity by a factor of $\sim2$ within $\sim3{\rm~yr}$, another characteristic of NLS1 galaxies. Short-term ($1000-100000{\rm~s}$) variability is not detected from RX~J1236.9+2656 due to poor signal-to-noise ratio of the $ROSAT$ data.

The soft X-ray spectrum of RX~J1236.9+2656 is steeper ($\Gamma_{X}\sim3.7$) 
than those of normal Seyfert 1s [$<\Gamma_{X}>$ ($90\%$ range)$=2.0-2.7$], and similar to those of NLS1 galaxies [$\Gamma_{X}(90\%{\rm~range})=2.3-3.7$]
(Grupe \etal 1998). Lack of intrinsic soft X-ray absorption over the Galactic 
value in RX~J1236.9+2656 is similar to the results found in normal Seyfert 1s and 
NLS1 galaxies. The steeper power-law index and blackbody model fit to the PSPC 
spectra of RX~J1236.9+2656 
indicate an ultra-soft nature of this object. The derived temperature of the blackbody, 
$kT\sim75{\rm~eV}$, is similar to those found in NLS1 galaxies (Brandt \& Boller 1998). 

The optical spectrum of RX~J1236.9+2656 appears to be typical of NLS1 galaxies. 
The FWHM of the H$\beta$ line (in the range of 1122--1392~km~s$^{-1}$) is narrower than 
those found in normal Seyfert 1s and is similar to those found in NLS1 
galaxies (FWHM$_{H\beta}\la2000{\rm~km~s^{-1}}$). The ratio 
$\frac{[O III]\lambda5007}{H\beta}$ ($\sim0.38$) for RX~J1236.9+2656 indicates 
that forbidden lines are weak, similar to that observed from NLS1 galaxies 
($\frac{[O III]\lambda5007}{H\beta}<3.0$). An Fe~II multiplet between $5070-5600{\rm~\AA}$ is also detected from RX~J1236.9+2656. At blue wavelengths, there is an indication of the presence of an Fe~II multiplet between $4435-4700{\rm~\AA}$ although our observation did not cover
 the multiplet fully. Thus the optical emission line parameters strongly 
suggest that RX~J1236.9+2656 is a NLS1 galaxy. Furthermore, the position of 
RX~J1236.9+2656 on the $\Gamma_{X}$--H$\beta$ line width plane (Fig. 8 of Boller \etal 1996) is consistent with other NLS1 galaxies.

\section{Conclusions}
A narrow-line Seyfert 1 galaxy, RX~J1236.9+2656, has been discovered based on the 
following soft X-ray and optical emission line properties: 
{\it (i) Steep soft X-ray spectrum ($\Gamma\sim3.7$), high soft X-ray luminosity ($1.6\times10^{43}{\rm~erg~s^{-1}}$), the lack of intrinsic soft X-ray absorption, and X-ray variability.
(ii) Narrow Balmer lines (FWHM$<2000{\rm~km~s^{-1}}$), weak [O III]$\lambda5007$ emission, and presence of Fe~II multiplets.}
\section{Acknowledgments}
This research has made use of $ROSAT$ archival data obtained through the
High Energy Astrophysics Science Archive Research Center, HEASARC,
Online Service, provided by the NASA-Goddard Space Flight Center. 
P.S and D.W. Hoard acknowledge support from NASA LTSA grant NAG 53345.
We thank an anonymous referee for his$/$her suggestions to reduce the size of this research note.

\end{document}